\documentclass[conference,10pt,letter]{IEEEtran}
\usepackage{graphicx}
\usepackage{amsmath}
\usepackage{amssymb}
\usepackage{math_symbols}
\usepackage{calc}
\usepackage{multirow}
\usepackage{mathrsfs}
\usepackage[english]{babel}
\usepackage{float}
\usepackage{cool}
\usepackage{url}
\usepackage{algorithm}
\IEEEoverridecommandlockouts

\usepackage{times}
\usepackage{url}
\usepackage{color,colortbl}
\usepackage{tikz}
\usepackage{caption}
\usepackage{subcaption}
\usetikzlibrary{calc,arrows,positioning,decorations,arrows.meta,shapes.symbols,decorations.pathreplacing}
\definecolor{TableCol}{rgb}{1,0.75,0}
\usepackage{enumitem}
\usepackage{tensor}
\usepackage{mathtools}
\usepackage{amsmath,amsfonts,amssymb,commath}
\usepackage{acronym}
\acrodef{kf}[KF]{Kalman filter} 
\acrodef{pf}[PF]{particle filter} 
\acrodef{kg}[KG]{Kalman gain}
\acrodef{ekf}[EKF]{extended \ac{kf}}
\acrodef{ckf}[CKF]{cubature \ac{kf}}
\acrodef{ukf}[UKF]{unscented \ac{kf}}
\acrodef{ai}[AI]{artificial intelligence} 
\acrodef{dnn}[DNN]{deep neural network} 
\acrodef{cnn}[CNN]{convolutional neural network} 
\acrodef{nn}[NN]{neural network}  
\acrodef{gnn}[GNN]{graph neural network} 
\acrodef{rnn}[RNN]{recurrent neural network} 
\acrodef{fc}[FC]{fully connected} 
\acrodef{mse}[MSE]{mean-squared error}
\acrodef{mc}[MC]{Monte Carlo}
\acrodef{snr}[SNR]{signal-to-noise ratio}
\acrodef{ss}[SS]{state-space}
\acrodef{ml}[ML]{machine learning}
\acrodef{gru}[GRU]{gated recurrent unit} 
\acrodef{lstm}[LSTM]{long short-term memory} 
\acrodef{ssm}[SSM]{state-space model}
\acrodef{rkn}[RKN]{recurrent Kalman network}
\acrodef{pbm}[PBM]{physics-based model}
\acrodef{apbm}[APBM]{data-augmented physics-based model}
\acrodef{tm}[TM]{true model}
\acrodef{rts}[RTS]{Rauch-Tung-Striebel}
\acrodef{ddm}[DDM]{data-driven model}
\acrodef{vae}[VAE]{variational autoencoder}
\acrodef{armax}[ARMAX]{autoregressive moving-average model}
\acrodef{pinn}[PINN]{physics-informed neural network}
\acrodef{narma}[NARMA]{nonlinear autoregressive moving-average}
\acrodef{nat}[NaT]{navigation and tracking}

\usepackage{siunitx}
\usepackage{bm}

\usepackage{times}
\usepackage{amssymb}
\usepackage{url}

\usepackage{cite} 
\usepackage{stfloats}  

\newcounter{steps}
	{\end{list}}

\newcommand{\APBMpar}{\bftheta}
\newcommand{\apbm}{\text{\tiny\ac{apbm}}}
\newcommand{\pbm}{\text{\tiny\ac{pbm}}}
\newcommand{\true}{\text{\tiny{true}}}
\newcommand{\data}{\text{\tiny{Data}}}

\newcommand{\con}{\text{\tiny{con}}}

\newcommand{\cb}[1]{\bm{#1}}

\begin{document}

\title{Interpretable Augmented Physics-Based Model for Estimation and Tracking}
	
\author{\IEEEauthorblockN{Ond\v{r}ej Straka, Jind\v{r}ich Dun\'{i}k}
	\IEEEauthorblockA{University of West Bohemia in Pilsen,\\
		Univerzitn\'{i} 8, 306 14 Pilsen, Czech Republic\\ Email: \{straka30, dunikj\}@kky.zcu.cz}
	\and
	\IEEEauthorblockN{Pau Closas}
	\IEEEauthorblockA{Northeastern University,\\ Boston, MA 02115, USA\\ Email: closas@ece.neu.edu}
    \and
	\IEEEauthorblockN{Tales Imbiriba}
	\IEEEauthorblockA{University of Massachusetts Boston,\\ Boston, MA, USA \\ Email: tales.imbiriba@umb.edu}
    \thanks{This work has been partially supported by the National Science Foundation under Awards ECCS-1845833 and CCF-2326559, and by the Ministry of Education, Youth and Sports of the Czech Republic under project ROBOPROX - Robotics and Advanced Industrial Production
CZ.02.01.01/00/22\_008/0004590.}
}

	\maketitle
	
	\selectlanguage{english}
	\begin{abstract}
State-space estimation and tracking rely on accurate dynamical models to perform well. However, obtaining an accurate dynamical model for complex scenarios or adapting to changes in the system poses challenges to the estimation process. Recently, augmented physics-based models (APBMs) appear as an appealing strategy to cope with these challenges where the composition of a small and adaptive neural network with known physics-based models (PBM) is learned on the fly following an augmented state-space estimation approach. A major issue when introducing data-driven components in such a scenario is the danger of compromising the meaning (or interpretability) of estimated states. In this work, we propose a novel constrained estimation strategy that constrains the APBM dynamics close to the PBM. The novel state-space constrained approach leads to more flexible ways to impose constraints than the traditional APBM approach. Our experiments with a radar-tracking scenario demonstrate different aspects of the proposed approach and the trade-offs inherent in the imposed constraints.
	\end{abstract}


\section{Introduction}\label{sec:intro}

Control and state estimation are fundamental to various applications such as navigation, space exploration, autonomous transportation, power systems, and robotics \cite{Dardari15,dehghanpour2018survey,haykin2004kalman,Vila17c,dunik2020state}. The tasks require mathematical models of the involved real systems.  The models incorporate latent states governed by, possibly sophisticated, dynamical models and observed through measurements obtained by various sensors. The sensors then determine the measurement model. 

Assuming that the measurement model is known or can be reliably obtained, classical state estimation approaches~\cite{BarShalom01, Sarkka13book} rely on the design of appropriate \ac{pbm} of the state dynamics.
Despite the extensive research in the past seven decades, obtaining the dynamical model still poses many challenges, especially related to designing or identifying its deterministic component, identifying its stochastic component (noise statistical properties), and the adaptation of such models 
to new operating conditions. 

Recent efforts have attempted to overcome parts of these challenges by leveraging data-driven \ac{ml} or, more generally, \ac{ai} algorithms. AI-based modeling of state dynamics can be roughly classified in terms of the \ac{ss} components that the data-driven model characterizes or is embedded in.
In general \ac{ss} structures, the whole model of state dynamics is identified from data and replaced by data-driven models such as an~\ac{nn} or \ac{gp}.
In~\cite{suykens1995a}, \ac{fc} \acp{nn} were used to represent the function in state dynamics,
and in \cite{masti2021a}, the authors use predictive partial autoencoders to find the state and its mapping to the measurement.
In~\cite{gorji2008a}, \ac{fc} \acp{nn} trained by the expectation-maximization algorithm were used to represent both functions while~\cite{wang2017a} utilized \ac{lstm} \acp{nn} for this purpose.

A comparison of the performance of several general \ac{nn}-based \ac{ss} structures was presented in~\cite{gedon2021a}, considering \ac{lstm}, \ac{gru}, and \acp{vae}.
The authors assumed Gaussian likelihood and combined the \ac{rnn} and \ac{vae} into a deep-state space model.
Stochastic \acp{rnn} were introduced for system identification in~\cite{fraccaro2016a} to account for stochastic properties of the \ac{ss} model. A similar approach is followed by \ac{pinn}~\cite{arnold2021} modeling the state equation by a \ac{nn} which training is constrained by the physics describing the \ac{pbm}.
The limitations of those, however, include poor explainability of the resulting model or the usual need for supervised learning, i.e., true state availability. Although there is a recent focus on \ac{nn}-based approaches, abundant literature exists on \ac{gp}-based approaches for \ac{ss} models~\cite{lydeard2025integrated, Hamelijnck_GP_neurips24, sarkka2021use}. For instance, in~\cite{sarkka2021use} the authors propose a \ac{gp}-based strategy for system identification, while physics-informed \ac{gp} models were considered in \cite{Hamelijnck_GP_neurips24} where a variational \ac{gp} strategy was used. In~\cite{lydeard2025integrated}, a linear physical-model is leveraged as the mean of \ac{gp} formulation of the state dynamics. Despite the rich literature, in this work we will focus our discussions on \ac{nn}-based strategies.

As an alternative to designing a model for the estimation, \ac{ml} can be used directly in state estimation to compensate for the limited model quality.
Here, \ac{ml} has found its place largely in the context of target tracking as the tracked object dynamic is typically unknown, and the estimators employ approximations such as nearly constant velocity, nearly constant acceleration, or coordinated-turn models. 
In this context, \ac{ml} approaches can be classified into \textit{(i)} hybrid-driven (combining standard Bayesian estimation with \acp{nn}) and \textit{(ii)} purely data-driven approaches.
The hybrid-driven approaches either provide corrections to estimates calculated by standard state estimation methods, e.g.,~\cite{vaidehi1999a} or~\cite{shaukat2021a}, represent the filter gain by a \ac{nn} in KalmanNet~\cite{revach2022kalmannet}  or directly estimate the state, e.g.,~\cite{gao2019a}, \cite{jung2020a}, \cite{ghosh2023danse}. 
Purely data-driven approaches use plain data to learn mappings from observations to the states to avoid complications of the hybrid-driven methods,~\cite{becker2019recurrent},~\cite{zhai2019a}.

A significant disadvantage of these state estimation approaches is that the \ac{nn} is trained for a specific model, data, sensor quality, choice of sensors, and estimator type.
Any change of these requires retraining the \ac{nn} as possibilities of transfer learning in the context of data-driven approaches to state estimation have only recently been investigated~\cite{imbiriba2022hybrid,imbiriba2023augmented,TangAPBMHOM2024, shlezinger2024ai}.

In this context, the hybrid adaptive frameworks that augment the \ac{pbm} with data-driven components become appealing for state estimation problems.  Introduced first in~\cite{imbiriba2022hybrid} and extended later in~\cite{imbiriba2023augmented,TangAPBMHOM2024,DuStKoTaImCl:24}, \ac{apbm}, have significant advantages compared to the previously discussed approaches regarding adaptability to new operation conditions, achieved through a joint state and parameter estimation approach, and interpretability of estimated states. The \ac{apbm} is a hybrid model augmenting the assumed \ac{pbm} dynamics with \acp{nn} providing extra flexibility.  We highlight that the flexibility of data-driven models allows one to model complex relationships between data and states. 


The interpretability, achieved by limiting the data-driven model contribution through control, can be viewed as an augmented likelihood model~\cite{imbiriba2022hybrid}.
Although effective, controlling the contribution of the data-driven component in the parameter space acts as a surrogate to the real objective of regularizing the \ac{apbm} transition function to lie in a neighborhood of the \ac{pbm} transition function.

In this paper, we propose a new control strategy for \acp{apbm} such that at every time step, the \ac{apbm} dynamics lies within an $\varepsilon$-neighborhood of the \ac{pbm}.
Such requirement is imposed in the \ac{ss} domain, leading to interpretable augmentation control in the sense that \emph{(i)}~the augmentation contribution to the \ac{pbm} can readily be quantified in physical units (of the state) and \emph{(ii)} the control can be enforced to some elements of the state only (e.g., the velocity components in the position-velocity-acceleration models).

The constraint is imposed by designing a constrained optimization problem. To solve the problem within a recursive Bayesian framework, we propose a projection strategy to map the main mass of the state distribution into the feasible area. An added advantage of the proposed approach is the ability to constrain only parts of the states, a feature that has not existed in previous works.
Experimental results with radar-based tracking systems demonstrate the efficacy of the proposed approach and the inherent trade-offs between fitting of measured states and the strength of the constraint. 

The remainder of this document is organized as follows. Section~\ref{sec:apbm} presents the \ac{apbm} and discusses the joint estimation of parameters and states. Section~\ref{sec:joint_est} discusses the proposed \ac{ss} constraint-based control strategy, while Section~\ref{sec:constraint_sol} discusses the projection strategy to enforce the constraints. Section~\ref{sec:exp} presents our experimental results and major findings. Finally, we present our concluding remarks in Section~\ref{sec:conc}.

\section{Augmented Physics-Based Model}\label{sec:apbm}
Consider the underlying $\mathrm{true}$ model, i.e., the data generator written as:
\begin{flalign}
    &\calM^{\true}:& \bfx_{k}&=\bff_{k-1}^{\,\true}(\bfx_{k-1},\bfu_{k-1})+\bfq^{\true}_{k-1},&&\label{eq:ssm1x}\\
    &&\bfy_{k}&=\bfh_k(\bfx_k,\bfr_k),&&\label{eq:ssm1z} 
\end{flalign}
where $\bfx_k\in\bbR^{d_x}$ is the state vector at time instant $t_k$, $\bfu_k$ is a control input vector, $\bfy_k$ is the available measurement, $\bff_k^{\,\true}$ is unknown, possibly complex and time-varying, state transition vector function and $\bfh_k$ is known measurement vector-valued function, $\bfq^\true_{k-1}, \bfr_k$ are the state and measurement noises with PDFs $\bfq_{k-1}^\true \sim \calN(\mathbf{0}, \bfQ), \bfr_k\sim\calN(\bfnul, \bfR)$, respectively.
As the measurement model~\eqref{eq:ssm1z} is assumed to be known, the rest of the exposition is focused on the state transition.

The \ac{pbm} obtained from first principles or classical system identification is described by the system dynamics:
\begin{flalign}
    &\calM^\pbm:&\bfx_{k}&=\bff_{k-1}^{\,\pbm}(\bfx_{k-1},\bfu_{k-1})+\bfq_{k-1}^\pbm,&&\label{eq:ssm2x}
\end{flalign}
\noindent where $\bff_{k-1}^{\,\pbm}$ is the physics-based dynamical model, and $\bfq_{k-1}^\pbm\sim \calN(\mathbf{0}, \bfQ^\pbm)$ is the additive state noise term.
Compared to~\eqref{eq:ssm1x}, its quality is limited due to insufficient knowledge of~\eqref{eq:ssm1x} or the pursuit of a simple model.

The \ac{apbm} dynamics augmenting~\eqref{eq:ssm2x} with a data-component can be written as:
\begin{flalign}
&\calM^\apbm\!\!:&\hspace{-0.3cm}\bfx_{k}&=\bff_{k-1}^\apbm\left(\bfx_{k-1}, \bfu_{k-1}, \bfd_{k-1};\APBMpar\right)+\bfq^{\apbm}_{k-1}&&\nonumber\\
&&&=\bfg\left(\bff_{k-1}^\pbm(\bfx_{k-1},\bfu_{k-1}),\bfx_{k-1}, \bfu_{k-1}, \bfd_{k-1};\APBMpar\right)\nonumber\\&&&+\bfq^{\apbm}_{k-1},&& \label{eq:ssm4x}
\end{flalign}
where the dynamics is constructed based on the composition of a data-driven function $\bfg(\cdot)$, parameterized by $\APBMpar$, and the PBM model $\bff_{k-1}^\pbm(\cdot)$. Additionally, the data-driven component may consider additional information encoded in the vector $\bfd_{k-1}$. For convenience, the additional information vector will be omitted in the sequel.

\subsection*{\ac{apbm} in Estimation and Tracking}
The key component of the tracking application is the estimation of the state $\bfx_k$ given a set of measurements $\bfy_{1:k}\coloneqq\{\bfy_1,\ldots,\bfy_k\}$. 
Since the \ac{apbm} involves the unknown \ac{nn} parameter $\APBMpar\in\bbR^{d_\theta}$, it must be estimated simultaneously with the state $\bfx_k$.
The joint state and parameter estimation problem becomes:
\begin{align}\label{eq:control_prop_orig}
    (\hat{\bfx}_{1:k}, \hat{\APBMpar}) &= \mathop{\arg\min}_{(\bfx_{1:k},\APBMpar)} \calC^\data (\bfy_{1:k},\bar{\bfy}_{1:k}) + \eta \calR(\bfx_{1:k},\bftheta),
\end{align}
where $\cal{C}^\mathrm{Data}$ and $\cal{R}$ are cost and regularization functionals, respectively, ${\bar{\bfy}}_{k}\triangleq \bfh_k(\bfx_k)$ such that $\bar{\bfy}_{1:k}\coloneqq\{\bar{\bfy}_{1},\ldots,\bar{\bfy}_{k}\}$, and $\eta\in\bbR_+$ is a regularization parameter governing the trade-off between model fit and regularization. 

In~\eqref{eq:control_prop_orig}, we assume stationary data-driven augmentation where $\APBMpar$ is assumed constant over time. A more interesting scenario considers time-varying (i.e., nonstationary) system dynamics that can benefit from time-varying data-driven models. This can be achieved by adapting the model parameters $\APBMpar$ over time~\cite{imbiriba2022hybrid, TangAPBMHOM2024}.

The recursive solution to the unconstrained state and parameter estimation problem can be obtained by leveraging Bayesian filtering techniques such as  \ac{kf} for linear Gaussian models, extended/sigma-points-based Kalman filters when considering nonlinear Gaussian models, or particle filters for the more general non-Gaussian scenarios. Here, we consider a nonlinear Gaussian assumption where many efficient approaches, such as the \ac{ckf}, can be leveraged. 

Following the joint estimation strategy in~\cite{imbiriba2022hybrid}, we augment the \ac{ss} with the \ac{apbm} parameters $\APBMpar$, with the following dynamical model: $\APBMpar_k = \APBMpar_{k-1} + \bfq^{\APBMpar}_{k-1}$. Thus, the solution of the joint, unconstrained optimization problem can be achieved by the augmented states estimates $\hat{\bfs}_k=[\hat{\APBMpar}_k, \hat{\bfx}_k]$ provided by the \ac{ckf}.

\section{Joint State-Parameter Estimation with Augmentation Control}\label{sec:joint_est}
The joint recursive state-parameter estimation problem can be formulated as a regularized optimization problem that can be solved recursively at each time step. A relevant point when embedding \ac{nn} models into the system dynamics is the interpretability of estimated states. This is paramount for safety-critical applications such as \cite{DO229D:06}.
In our previous works~\cite{imbiriba2022hybrid, imbiriba2023augmented, TangAPBMHOM2024}, we controlled the contribution of the \ac{nn} augmentation of \ac{apbm}s by directly regularizing the parameters $\APBMpar$ of the \ac{apbm} model.
Despite the good performance and the relative explainability of the estimated states, the regularization procedure used in~\cite{imbiriba2022hybrid,imbiriba2023augmented,TangAPBMHOM2024}, consists in constraining the APBM parameters, $\APBMpar_k$, to a neighborhood of $\bar{\APBMpar}$, where $\bar{\APBMpar}$ is the parameter value such that
\begin{align}
\bff^\apbm_k(\cdot;\APBMpar_k=\bar{\APBMpar})=\bfg\left(\bff^\pbm_k(\cdot),\cdot\,;\APBMpar_k=\bar{\APBMpar}\right)=\bff^\pbm_k(\cdot). 
\end{align}
Although effective, regularizing the contribution of the data-driven component in the parameter space acts as a surrogate to the real objective of regularizing the \ac{apbm} transition function to lie in a neighborhood of $\bff^\pbm_k$. 

Thus, in this paper, we propose different augmentation control metrics $\rho^\mathrm{SS}(\cdot,\cdot):\mathcal{X}\times \mathcal{X}\to \mathbb{R}$, that operate on the state space $\mathcal{X}\subset\mathbb{R}^d$. 
Here, we define two versions of augmentation control metrics. A relative version is
\begin{align}\label{eq:roh_ssr}
    \rho^\mathrm{SSR}(\bff^\apbm_k, \bff^\pbm_k) = \frac{\|\bff^\apbm_k(\hbfx_k,\bfu_k;\APBMpar_k) - \bff^\pbm_k(\hbfx_k,\bfu_k)\|_{\bfSigma}}{\|\bff^\pbm_k(\hbfx_k,\bfu_k)\|_{\bfSigma}}
\end{align}
and an absolute one is
\begin{align}\label{eq:roh_ssa}
    \rho^\mathrm{SSA}(\bff^\apbm_k, \bff^\pbm_k) = \|\bff^\apbm_k(\hbfx_k,\bfu_k;\APBMpar_k) - \bff^\pbm_k(\hbfx_k,\bfu_k)\|_{\bfSigma},
\end{align}
where $\hbfx_k$ represents an estimate of $\bfx_k$, and $\|\bfx\|_{\bfSigma} = \bfx\T\bfSigma\bfx$ with the weight $\bfSigma=(\bfQ^\pbm)^{-1}$ is used.

Using the weight $\bfSigma=(\bfQ^\pbm)^{-1}$ can be justified as follows: $\bfq^\pbm_k$ consists of two components: $\bfq^\true_k$ and a component overbounding the discrepancy between $\bff^\true_k$ and $\bff^\pbm_k$ \cite{DuStKoTaImCl:24}. Since the variance of 
 $\bfq^\true_k$ is unknown, the covariance $\bfQ^\pbm$ of $\bfq^\pbm_k$ will be used.

 The state and parameter estimation problem becomes:
\begin{align}\label{eq:control_prop}
    (\hat{\bfx}_{1:k}, \hat{\APBMpar}_{1:k}) &= \mathop{\arg\min}_{(\bfx_{1:k},\APBMpar_{1:k})} \calC^\data (\bfy_{1:k},\bar{\bfy}_{1:k}) + \eta \calR(\bfx_{1:k},\APBMpar_{1:k})\nonumber\\
     &\mathrm{s.t.} \quad \rho^\mathrm{SS}(\bff^\apbm_k, \bff^\pbm_k) \leq \epsilon, \vspace*{-.3cm}
\end{align}
where the parameter $\epsilon \in\bbR_+$ defines a closed ball of feasible solutions centered on $\bff^\pbm_k$ and $\rho^\mathrm{SS}$ is an augmentation control metrics \eqref{eq:roh_ssr} or \eqref{eq:roh_ssa}.  

The optimization problem in~\eqref{eq:control_prop} confines the predicted states to lie in a ball around the PBM prediction (see Figure~\ref{fig:illustration}). Another interesting aspect of the formulation is its easy application to specific parts of the state space, which allows for a flexible method that can be used to constrain only the unmeasured components of the states. 
The solution to the constrained joint state-parameter estimation~\eqref{eq:control_prop} is described in the following section.


\section{Estimation with constraint in state space}\label{sec:constraint_sol}
Imposing the augmentation control constraints into the \ac{ckf} solution can be challenging, especially when considering nonlinear inequality constraints such as the ones defined in Eqs.~\eqref{eq:control_prop} with~\eqref{eq:roh_ssr} and~\eqref{eq:roh_ssa}.
 Due to the nonlinearity of the constraint
 \begin{align}\label{eq:ineq_constraint}
 \rho^\mathrm{SS}(\bff^\apbm_k(\hbfx_k,\bfu_k;\APBMpar_k), \bff^\pbm_k(\hbfx_k,\bfu_k)) \leq \epsilon
 \end{align}
 and the need to obtain a covariance of the constrained estimate,
 the projection approach will be used.

 The constraint~\eqref{eq:ineq_constraint} defines implicitly a set $\APBMpar_k\in\Theta(\hbfx_k,\bfu_k)$ such that for a given $\bfx_k$ and  $\bfu_k$ it holds
 \begin{multline}
     \APBMpar_k\in\Theta(\hbfx_k,\bfu_k) \\ \implies  \rho^\mathrm{SS}(\bff^\apbm_k(\hbfx_k,\bfu_k;\APBMpar_k), \bff^\pbm_k(\hbfx_k,\bfu_k)) \leq \epsilon.
 \end{multline}
The proposed constrained estimation checks the validity of the constraint for the current estimates of the state $\hbfx_k$ and the parameter $\hat{\APBMpar}_k$. If the constraint~\eqref{eq:ineq_constraint} is not satisfied, the parameter $\hat{\APBMpar}_k$ is shifted towards $\bar{\APBMpar}$ (used in augmentation control in the parameter space\footnote{
    $\bar{\APBMpar}$ is defined such that
         $\bff^\apbm_k(\hbfx_k,\bfu_k;\bar{\APBMpar})= \bff^\pbm_k(\hbfx_k,\bfu_k),\, \forall\hbfx_k,\forall\bfu_k$.
}~\cite{imbiriba2022hybrid}), which lies in the region $\bar{\APBMpar}\in\Theta(\hbfx_k,\bfu_k)$ so that~\eqref{eq:ineq_constraint} is satisfied.

In particular, the projected parameter denoted as $\hat{\APBMpar}_k(\kappa)$ parametrized by a projection parameter $\kappa\in[0,1]$ is given by
\begin{align}\label{eq:APBM_kappa}
\hat{\APBMpar}_k(\kappa) = \kappa\cdot\hat{\APBMpar}_{k}+(1-\kappa)\cdot\bar{\APBMpar},
\end{align}
such that $\hat{\APBMpar}_k(1)=\hat{\APBMpar}_k$ and $\hat{\APBMpar}_k(0)=\bar{\APBMpar}$.
Then, a maximization over $\kappa$ with a nonlinear equality constraint is performed
\begin{align}\label{eq:kappa}
\kappa^* &= \max_{\kappa\in]0,1]} \kappa,\\\nonumber
\text{s.t. }&\rho^\mathrm{SS}\left(\bff^\apbm_k(\hbfx_k,\bfu_k;\hat{\APBMpar}_k(\kappa)), \bff^\pbm_k(\hbfx_k,\bfu_k)\right)=\epsilon.
\end{align}
The value $\hat{\APBMpar}_k(\kappa^*)$ then satisfies $\hat{\APBMpar}_k(\kappa^*)\in\bfTheta(\hbfx_k,\bfu_k)$.


For the constrained estimation, the uncertainty of the estimate  $\hat{\bfs}_k=[\hat{\bfx}_k,\hat{\APBMpar}_k]$ expressed through the covariance matrix
\begin{align}\label{eq:pss}
\bfP_{k}^{ss} = 
\begin{bsmallmatrix}
\bfP_{k}^{\bfx\bfx}&   \bfP_{k}^{\bfx\APBMpar}\\
\bfP_{k}^{\APBMpar\bfx}& \bfP_{k}^{\APBMpar\APBMpar}
\end{bsmallmatrix}
\end{align}
shall be taken into account. 
Since the state is estimated by the \ac{ckf}, which uses cubature points in the estimation, the uncertainty associated with the estimate is respected in the constraining by following the procedure given by \eqref{eq:APBM_kappa} and \eqref{eq:kappa} for all $n_S = 2(d_x+d_{\APBMpar})$ cubature points $\calS_{k,i}=[\calS^x_{k,i},\calS^{\APBMpar}_{k,i}],\, i=1,\ldots,n_S$ associated with $\hbfs_k$ and $\bfP_{k}^{ss}$.
Then, $n_S$ optimizations for each 
$\calS_{k,i}$ lead to $\kappa^*_i,\, i=1,\ldots n_S$ s.t.
\begin{align}
    \calS^{\APBMpar}_{k,i}(\kappa^*_i)\in\bfTheta(\calS^x_{k,i},\bfu_k).
\end{align}
Then, the minimum 
\begin{align}
    \kappa^*_{\min} = \min_{j=1,\ldots n_S}\kappa^*_j
\end{align}
is used to construct constrained cubature points 
\begin{align}
    \calS^\con_{k,i}=[\calS^x_{k,i},\calS^{\APBMpar,\con}_{k,i}(\kappa^*_{\min})],\, i=1,\ldots,n_S.
\end{align}
These constrained cubature points propagated through the \ac{apbm} dynamics
\begin{align}
    \bff^\apbm_k(\calS^x_{k,i},\bfu_k;\calS^{\APBMpar,\con}_{k,i}(\kappa^*_{\min}))
\end{align}
are then used to construct the constrained predicted estimate $\hbfs_{k+1|k}^\con$ and associated covariance matrix $\bfP_{k+1|k}^{ss,\con}$, which respect the constraint~\eqref{eq:ineq_constraint}.

The procedure for calculating the constrained state and parameter estimate respecting the augmentation control through the constraint~\eqref{eq:ineq_constraint} is illustrated in Figure~\ref{fig:illustration}, and the proposed constrained state estimation is described in Algorithm~\ref{alg:algorithm}.
\begin{figure*}
    \centering    
    \begin{tikzpicture}[scale=0.8]
\definecolor{parspacecol}{RGB}{66,62,195}
\definecolor{statespacecol}{RGB}{195,66,62}
\definecolor{ellipsecol}{RGB}{62,195,66}
\colorlet{sigmalinecol}{ellipsecol!70!white}
\colorlet{sigmaarrowcol}{parspacecol!70!white}
\tikzstyle{sigmapoint}=[fill,inner sep=0pt,minimum size=2pt]
\draw[color=statespacecol,line width=1pt,fill=statespacecol!10!white]   (0,0) circle[radius=2.5cm];
\fill   (0,0) circle[radius=2pt]  node [above,font=\small,anchor=south west] (pbm)  {$\bff^{\pbm}_{k}(\hbfx_{k},\bfu_k)$};
\fill   (2cm,3cm) circle[radius=2pt]  node [above,font=\small,anchor=south west] (apbm)  {$\bff^{\apbm}_{k}(\hbfx_{k},\bfu_k;\hat{\APBMpar}_k)$};
\draw [decorate,decoration={brace,amplitude=5pt,mirror}]
  (-2.5cm,0) -- (0,0) node[midway,yshift=-1em]{$\epsilon$};
\draw[line width=1pt, -{Stealth[length=3mm]}] ($(apbm.south west)+(-1mm,1mm)$) to [out=150,in=80] ((0.5,2);
\begin{scope}[shift={(20mm,0mm)}]
\coordinate (thetabar) at (8cm,0);
\coordinate (theta) at (10cm,4cm);
\node[color=parspacecol,line width=1pt,fill=parspacecol!10!white] [cloud, cloud puffs=9, draw, minimum width=4cm, minimum height=5cm] at (87mm, 7mm) {};
\fill (thetabar) circle[radius=2pt]  node [above,font=\small,anchor=north west] {$\bar{\APBMpar}$};
\fill (theta) circle[radius=2pt]  node [above,font=\small,anchor=south east,shift={(1mm,0mm)}]  {$\hat{\APBMpar}_k$};
\draw[color=parspacecol,latex-,dashed] (100mm,-8mm) -- (110mm,-16mm) node[color=black,at end,anchor=north west] {$\Theta(\hbfx_k,\bfu_k)$};
\draw[rotate around={30:(theta)},color=ellipsecol] ($(theta)+(3mm,1mm)$)
arc [start angle=0,   end angle=360,  x radius=10pt,  y radius=20pt]
                                node[sigmapoint,pos=0] (n1) {} 
                                node[sigmapoint,pos=.25] (n2) {} 
                                node [sigmapoint,pos=.5] (n3) {}  
                                node [sigmapoint,pos=.75] (n4) {};

\coordinate (ellipseinside) at ($(theta)!0.30!(thetabar)$);
\fill[color=gray] (ellipseinside) circle[radius=2pt];
\draw[rotate around={30:(theta)},color=ellipsecol] ($(ellipseinside)+(2.8mm,1mm)$)
arc [start angle=0,   end angle=360,  x radius=8pt,  y radius=13pt]
                                node[sigmapoint,pos=0] (m1) {} 
                                node[sigmapoint,pos=.25] (m2) {} 
                                node [sigmapoint,pos=.5] (m3) {}  
                                node [sigmapoint,pos=.75] (m4) {};
\draw (thetabar) -- (theta);
\node[font=\small,shift={(18mm,0mm)}] (thetakappa) at (ellipseinside) {$\hat{\APBMpar}_k(\kappa^*)$};
\draw[-latex,dashed] (thetakappa) -- (ellipseinside);
\foreach \i in {1,2,...,4}{
\draw[draw=sigmalinecol,dashed] (thetabar) -- (n\i);
\draw[draw=black,-latex] (n\i) -- (m\i);
}
\node[font=\small] at (25mm,-28mm) {$\bff^{\pbm}_{k}(\hbfx_{k},\bfu_k)=\bff^{\apbm}_{k}(\hbfx_{k},\bfu_k;\bar{\APBMpar})$};
\end{scope}
\node[anchor=east,color=statespacecol] (ss) at (55mm,45mm) {state space $\calX$};
\node[anchor=west,color=parspacecol] (ss) at (55mm,45mm) {parameter space $\bbR^{d_\theta}$};
\draw[thick,lightgray,dashed,dash pattern={on 20pt off 4pt}] (55mm,45mm) -- ++(0,-70mm);
\end{tikzpicture}
    \caption{Illustration of the constrained parameter estimation. Left: the situation in state space illustrating the constraint, Right: the modification of the unconstrained $\hat{\APBMpar}_k$ to achieve the constrained $\hat{\APBMpar}_k(\kappa^*)$.}
    \label{fig:illustration}
\end{figure*}
\begin{algorithm}
\caption{Procedure of augmentation control in state space through constrained estimation}\label{alg:algorithm}
\begin{enumerate}
    \item \textbf{Assume:} Estimates $\hbfx_k$ of $\bfx_k$ and $\hat{\APBMpar}_k$ of $\APBMpar_k$. 
    \item \textbf{Checking constraint validity:} If for $\hbfx_{k}$, $\bfu_k$, and $\hat{\APBMpar}_k$ the constraint~\eqref{eq:ineq_constraint} is not satisfied,
    generate cubature points $\calS_{k,i}=[\calS^x_{k,i},\calS^{\APBMpar}_{k,i}],\, i=1,\ldots,n_S$ based on $\hbfs_k=[\hat{\bfx}_k,\hat{\APBMpar}_k]$ and $\bfP_k^{ss}$ \eqref{eq:pss}.
    \item \textbf{Optimizing the projection parameter:} For each cubature point, perform optimization by optimizing over~$\kappa$
    \begin{align*}
        \kappa^*_i &= \max_{\kappa_i\in]0,1]} \kappa_i\\
        \text{s.t.\! }&\rho^\mathrm{SS}\left(\!\bff^\apbm_k\!\left(\calS^x_{k,i},\bfu_k;\calS^{\APBMpar}_{k,i}(\kappa_i)\right), \bff^\pbm_k\!(\calS^x_{k,i},\bfu_k)\right)\!=\!\epsilon,
    \end{align*}
    where 
    \begin{align}
     \calS^{\APBMpar}_{k,i}(\kappa_i) = \kappa_i\cdot\calS^{\APBMpar}_{k,i}+(1-\kappa_i)\cdot\bar{\APBMpar}.
     \end{align}    
     \item \textbf{Constrained cubature points:} Use the minimum 
\begin{align}
    \kappa^*_{\min} = \min_{j=1,\ldots n_S}\kappa^*_j
\end{align}
 to construct constrained cubature points 
\begin{align}
    \calS^\con_{k,i}=[\calS^x_{k,i},\calS^{\APBMpar,\con}_{k,i}(\kappa^*_{\min})],\, i=1,\ldots,n_S.
\end{align}
\item \textbf{Time update:} Propagate the constrained cubature points through
\begin{align}
    \bff^\apbm_k\left(\calS^x_{k,i},\bfu_k;\calS^{\APBMpar,\con}_{k,i}(\kappa^*_{\min})\right)
\end{align}
and  use them to construct the predicted estimate $\hbfs^{\con}_{k+1|k}$ and associated covariance matrix $\bfP_{k+1|k}^{ss,\con}$, which respect the constraint~\eqref{eq:ineq_constraint}.
   \item \textbf{Measurement update:} Perform the standard measurement update of the \ac{ckf} to obtain the state and parameter estimate $\hat{\bfs}_{k+1}=[\hat{\APBMpar}_{k+1}, \hat{\bfx}_{k+1}]$ and associated  covariance matrix $\bfP_{k+1}^{ss}$. 
\end{enumerate}
\end{algorithm}

\section{Numerical Experiments}\label{sec:exp}

We test the proposed regularization approach using the two-dimensional synthetic radar target tracking scenario discussed in~\cite{imbiriba2023augmented}. The simulation considers a time-varying dynamics
additive constant velocity~\cite{closas2011particle} and sinusoidal~\cite{Arasaratnam09} terms:
\begin{align}\label{eq:target_state_eq}
 \mathcal{M}^\mathrm{true}:&& {\mathbf x}_k &= (\bfF +\bfG_{k-1}) {\mathbf x}_{k-1} + \bfM\, \cb{q}_{k-1} ~, \\
  &&\Omega_k &= \Omega_{k-1} + v_{k-1}, \nonumber
\end{align}
with 
$$
\bfF = \begin{psmallmatrix}
                          1 & T_s & 0 & 0 \\
                          0 & 1 & 0 & 0 \\
                          0 & 0 & 1 & T_s \\
                          0 & 0 & 0 & 1
                        \end{psmallmatrix}, \,\,
  \mathbf M = \begin{psmallmatrix}
                          T_s^2/2 & 0 \\
                          T_s & 0 \\
                          0 & T_s^2/2 \\
                          0 & T_s
                        \end{psmallmatrix},
  $$
  $$
  \, \bfG_{k-1} = \begin{psmallmatrix}
                          0 & \frac{\sin\Omega_{k-1} T_s}{T_s} & 0 & -\frac{1 - \cos \Omega_{k-1}T_s}{\Omega_{k-1}} \\
                          0 & \cos \Omega_{k-1}T_s & 0 & -\sin \Omega_{k-1}T_s \\
                          0 & \frac{1 - \cos \Omega_{k-1}T_s}{\Omega_{k-1}} & 0 & \frac{\sin \Omega_{k-1}T_s}{\Omega_{k-1}} \\
                          0 & \sin \Omega_{k-1}T_s & 0 &  \cos \Omega_{k-1}T_s
  \end{psmallmatrix},
  $$
\noindent ${\mathbf x}_k = (x_k, \dot{x}_k, y_k, \dot{y}_k)^\top$
being a state vector, composed of the two-dimensional position
(${\mathbf p}_k = (x_k, y_k)^\top$) and velocity  ($\dot{\mathbf p}_k = (\dot{x}_k, \dot{y}_k)^\top$) of the target, and $\Omega_k$ being an angle state. 
In our experiments, we generate trajectories with the following parameters: $T_s=1$s is the sampling period, and
$\bfq_k\sim \calN \left( {\mathbf 0}, 0.01 \cdot
{\mathbf I} \right)$ is the process noise that models the acceleration as a random term. $v_k\sim\calN(0, 10^{-5})$ is the process noise for the angle $\Omega_k$. The true trajectory was initialized
at $\bfx_0 = (50, 0, 50, 0)^\top$ and $\Omega_0=0.05\pi$.

Synthetic measurements from two collocated sensors measuring received signal strength (RSS) and bearings were considered:
\begin{equation}\label{eq:target_meas_eq}
{\mathbf y}_k =  \mathbf{h}(\mathbf p^\mathrm{s}_k, {\mathbf p}_k) +  {\mathbf r}_k  = \left( \begin{array}{c}
                         10 \log_{10} \left( \frac{\Psi_0}{\parallel {\mathbf p^\mathrm{s}_k} - {\mathbf p}_k \parallel^\alpha} \right) \\
                         \angle ({\mathbf p^\mathrm{s}_k}, {\mathbf p}_k )
                       \end{array}
\right) + {\mathbf r}_k ~,
\end{equation}
\noindent with ${\mathbf p^\mathrm{s}_k}$ being the position of the sensors, ${\mathbf p}_k$ the unknown position of the target,
$10 \log_{10} \left( \Psi_0 \right) = 30$ dBm, $\alpha=2.2$ the path loss
exponent, $\angle ({\mathbf p^\mathrm{s}_k}, {\mathbf p}_k )$ denoting the
angle between locations ${\mathbf p^\mathrm{s}_k}$ and ${\mathbf p}_k$ in radians, and
${\mathbf r}_k \sim {\mathcal N} \left( {\mathbf 0}, \bfR_k \right)$, $\bfR_k=\textrm{diag}(0.1 , 0.1),$ the measurement noise.

We consider an APBM model of the additive form:
\begin{align}\label{eq:target_state_eq_apbm}
\mathcal{M}^\apbm:\bfx_k &= \phi_0\bfF\bfx_{k-1} + \phi_1\gamma(\bfx_{k-1}; \cb{\omega}) + \bfM\, \cb{q}_{k-1}, \nonumber
\end{align}
\noindent where $g(\cdot)$ is the \ac{apbm} parameterized by $\cb{\theta}=\{\phi_0,\phi_1,\cb{\omega}\}$,
$\gamma(\cdot)$ is a \ac{nn} parameterized by $\cb{\omega}$ with one hidden layer with $5$ neurons and ReLu activation function, an output layer with $4$ neurons and linear activation, leading to a total of $42+2$ parameters including bias terms. 

The experiments analyzed the performance of \ac{pbm}, \ac{apbm} with augmentation control directly regularizing $\APBMpar$~\cite{imbiriba2022hybrid} referred to simply as `APBM', and the proposed \ac{apbm} controlling augmentation in the \ac{ss} with absolute metric~\eqref{eq:roh_ssa}  referred to as `APBM\_SSA\_e=$\epsilon$', where $\epsilon$ is the value of the threshold.  

All experiments were performed over $N_{MC}=100$ \ac{mc} runs, with data generated over $T=1000$ seconds (samples).
The state estimate quality was evaluated using three criteria: \emph{(i)} the root mean square error (RMSE), \emph{(ii)} the averaged normalized estimation error squared (ANEES) \cite{LiZha:06}, and \emph{(iii)} the cumulative distribution function (CDF) of the estimate errors. Whereas the RMSE and the error CDF evaluate the estimate error, i.e., the estimate accuracy,
the ANEES assesses the estimate consistency, i.e., the similarity of the estimate mean square error (MSE) and the provided estimate error covariance matrix.

The RMSE for a given time $k$ computed over all \ac{mc} realizations is given by:
\begin{equation}
    \text{RMSE}_k = \sqrt{\frac{\sum_{r=1}^{N_{MC}}\|\bfx^{(r)}_k-\hat{\bfx}^{(r)}_k\|^2}{dN_{MC}}}
\end{equation}
with $d$ is the dimension of $\bfx_k$, $N_{MC}$ is the number of \ac{mc} runs, and $\hat{\bfx}_k$ represents the estimated states.

The ANNES is given by~\cite{LiZha:06a}:
\begin{align}\label{eq:annes}
        \text{ANEES}_k =\ \tfrac{1}{N_{MC}}\sum_{r=1}^{N_{MC}}(\bfx_{k}^{(r)}-\hbfx_{k}^{(r)})^\top(\bfP_{k}^{(r)})^{-1}(\cdot),
\end{align}
where $\bfP_k^{(r)}$ is the state covariance matrix produced by the filter at time $k$ and \ac{mc} run $r$. The weighted error ANEES in~\eqref{eq:annes} should follow a $\chi^2_d$ distribution with $d$ degrees of freedom, and its average should be around $d$. Thus, ANNES close to $d$ indicates a credible estimator of the error in the sense that both the estimated mean $\hbfx_k$ (and its error), and covariance $\bfP_k$ are consistent. When ANEES is smaller than $d$, the estimator is pessimistic in the sense that the error covariance produced by the estimator is higher than the true error covariance. On the other hand, if ANNES is bigger than $d$, then the estimator is said to be optimistic, that is, the estimated error covariance is under-estimated. 

\begin{figure*}[htb]
    \centering
    \includegraphics[width=0.49\linewidth]{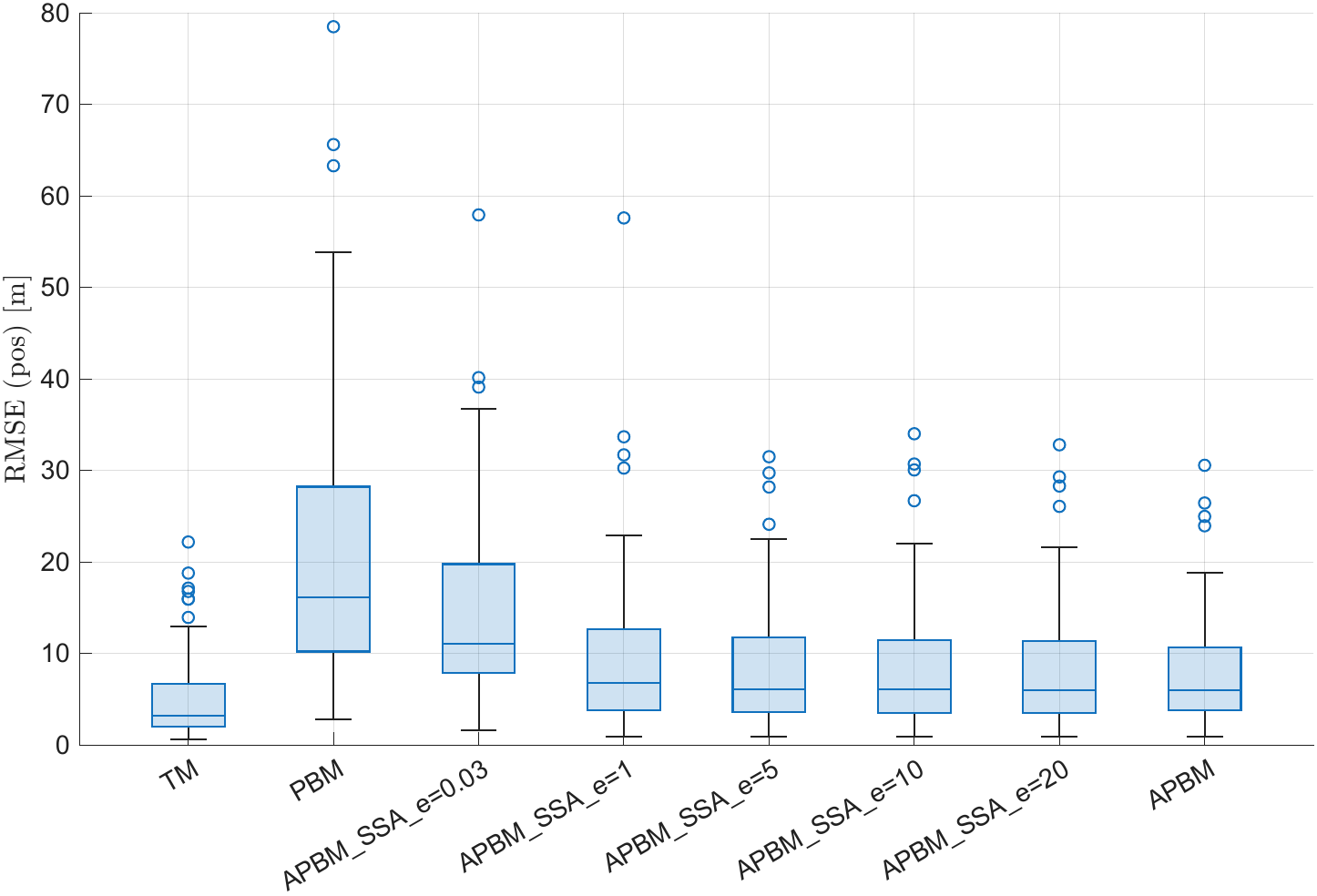}
    \includegraphics[width=0.49\linewidth]{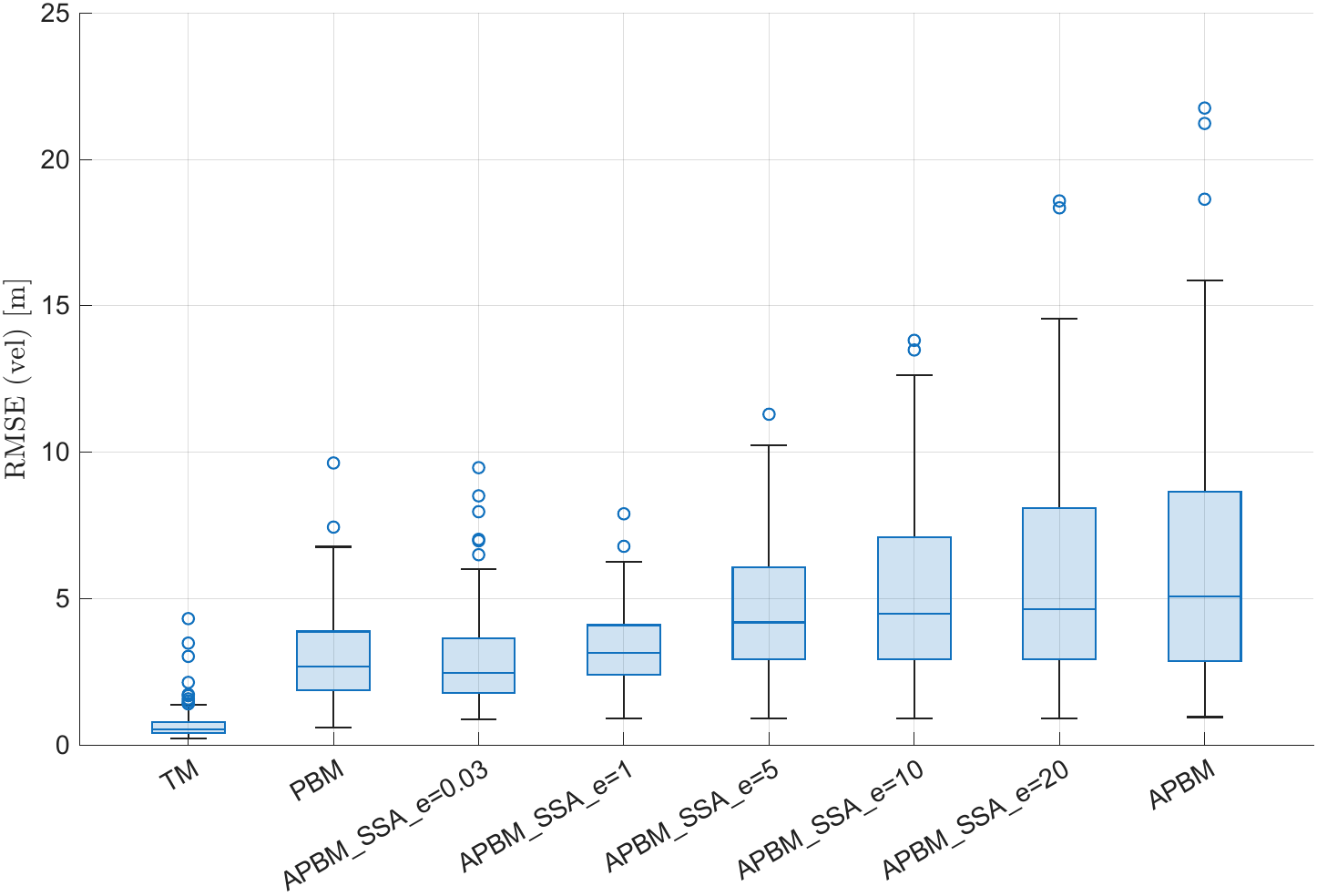}
    \includegraphics[width=0.49\linewidth]{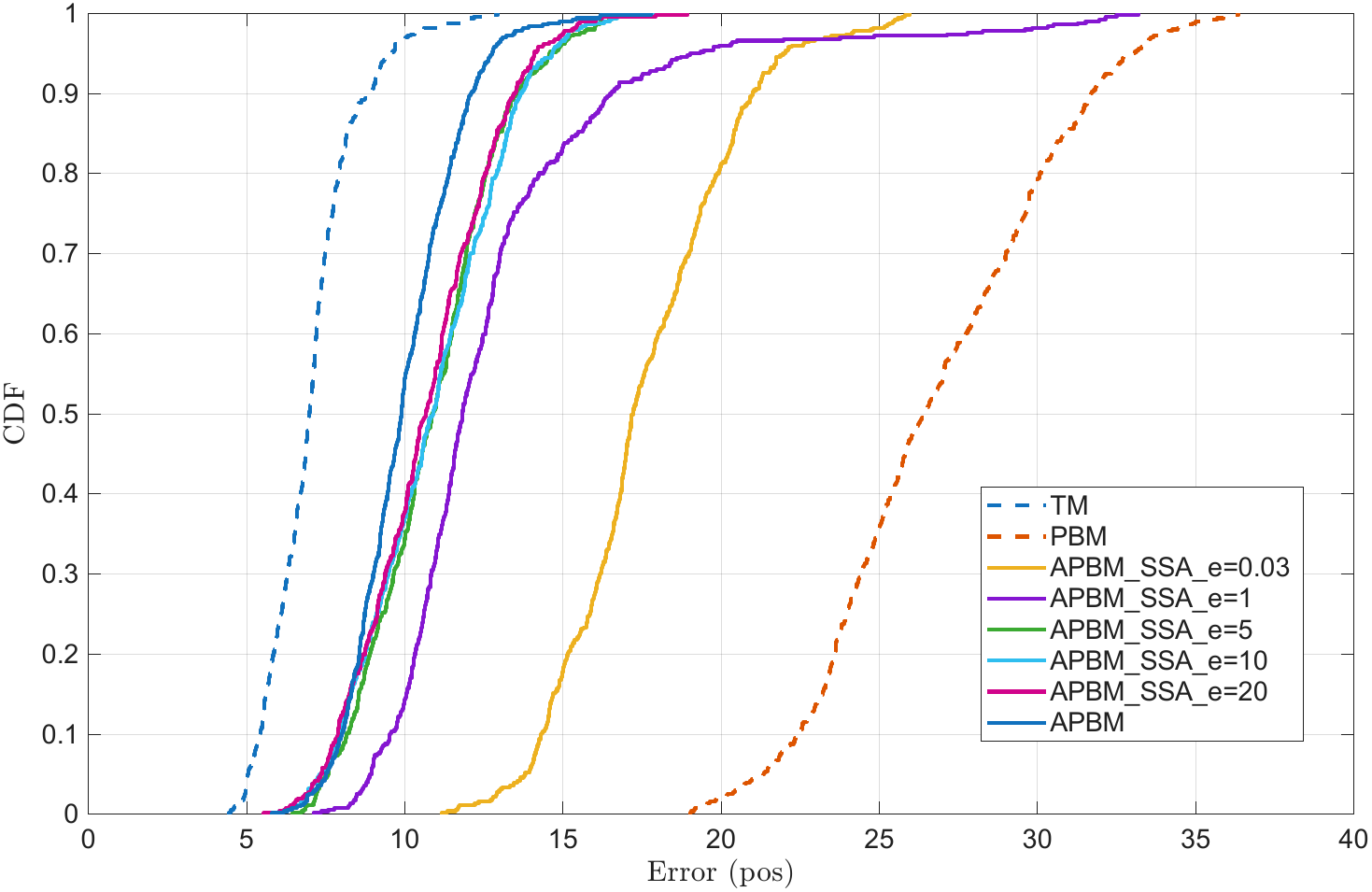}
    \includegraphics[width=0.49\linewidth]{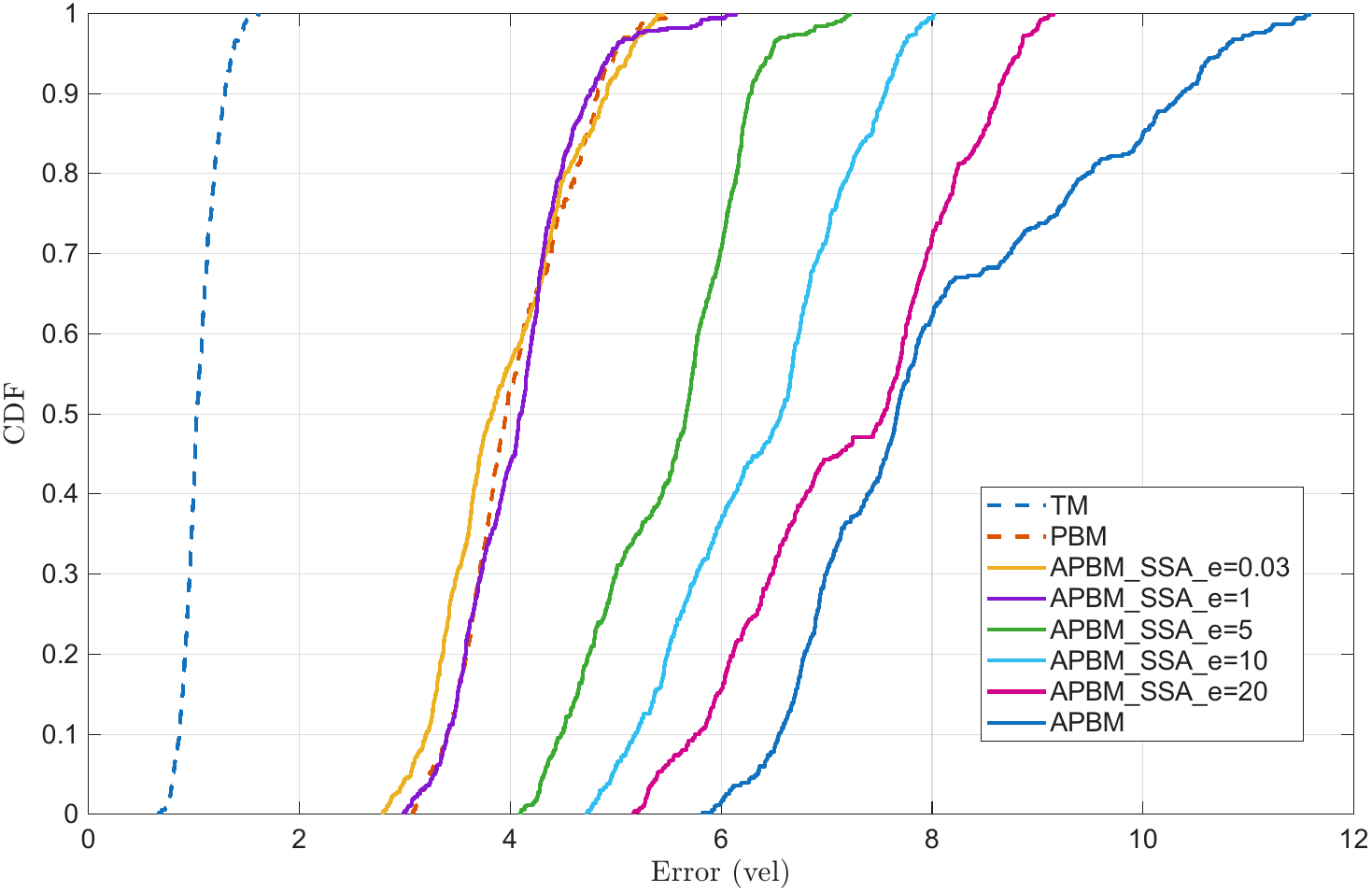}
    \caption{Results under fully-constrained states. (\textbf{top-left}): RMSE for position estimates; (\textbf{top-right}): RMSE for velocity estimates;  (\textbf{bottom-left}): CDF of the position error; (\textbf{bottom-right}): CDF of the velocity error.}
    \label{fig:results_fullyconstrained}
\end{figure*}

\begin{figure}
    \centering
    \includegraphics[width=0.99\linewidth]{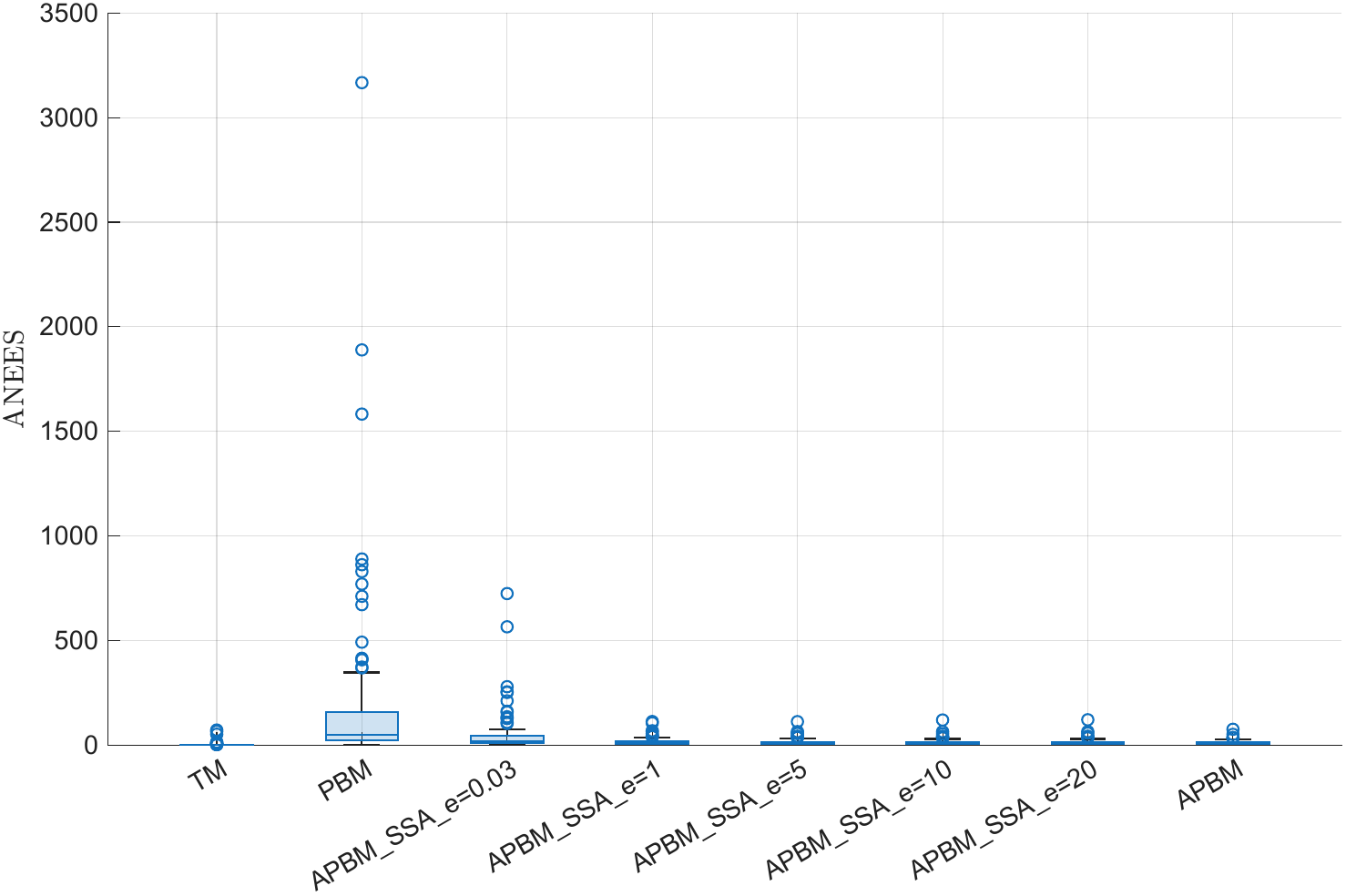}
    \caption{Results under fully-constrained states. ANEEs.}
    \label{fig:ANEEs_fullyconstrained}
\end{figure}

\begin{figure*}[htb]
    \centering
    \includegraphics[width=0.49\linewidth]{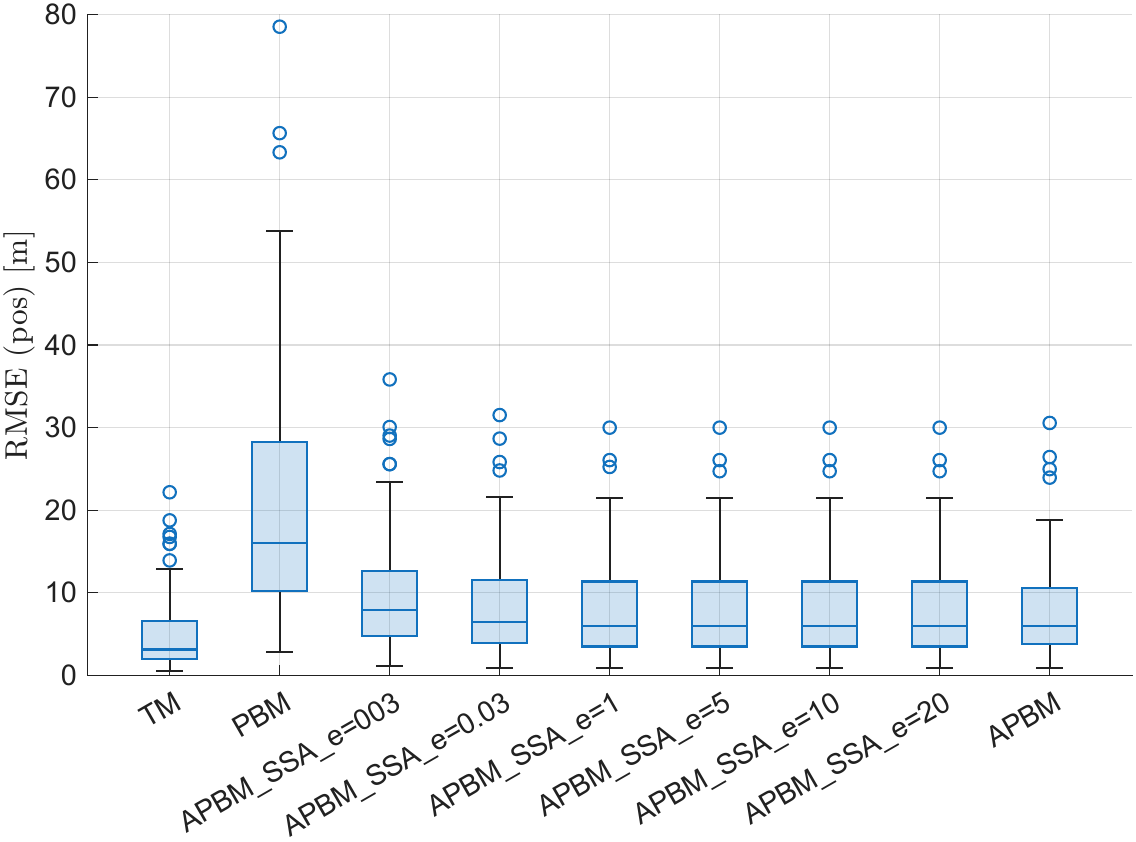}
    \includegraphics[width=0.49\linewidth]{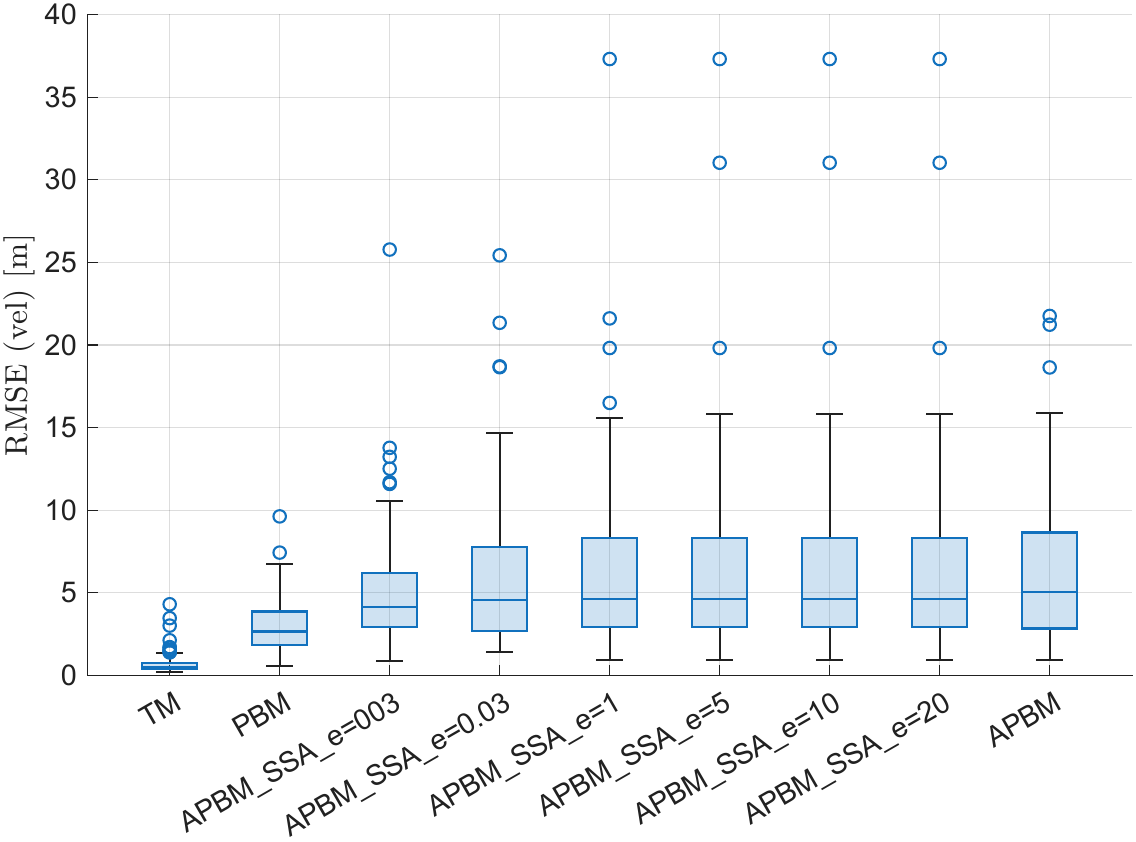}
    \includegraphics[width=0.49\linewidth]{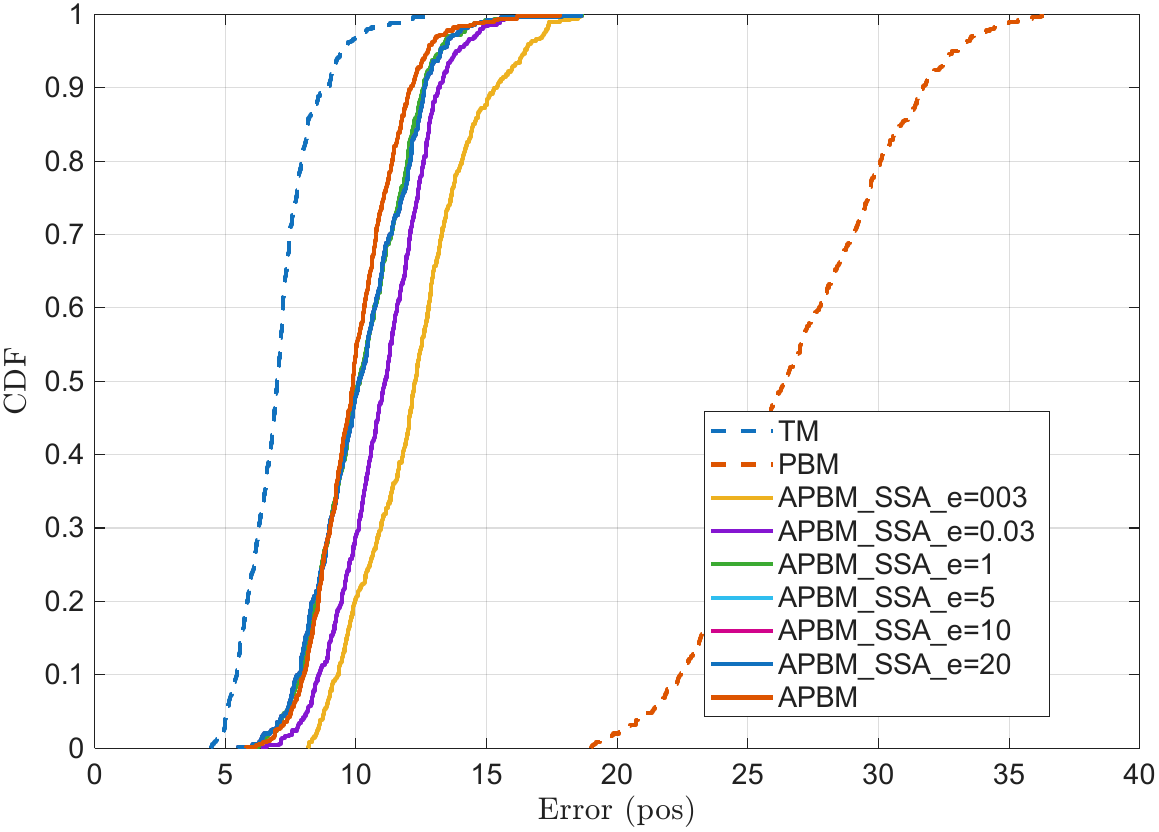}
    \includegraphics[width=0.49\linewidth]{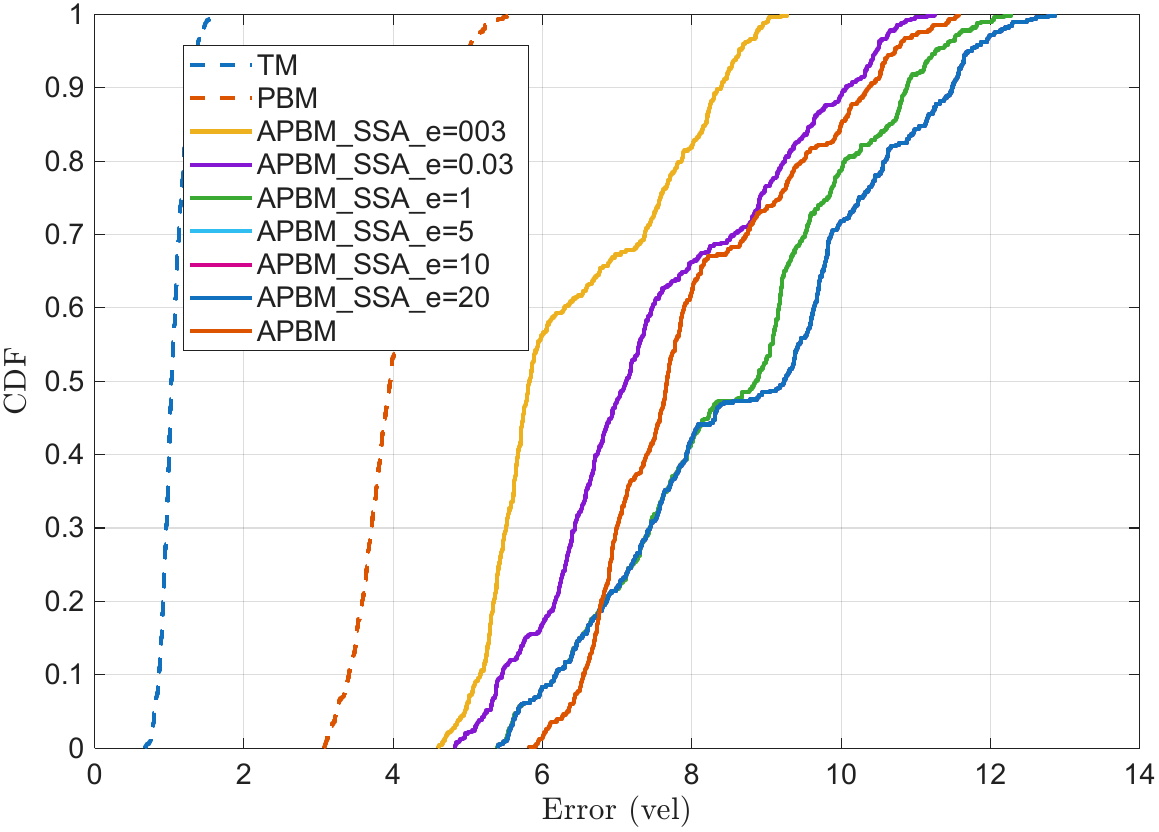}
    \caption{Results under constrained velocity. (\textbf{top-left}): RMSE for position estimates; (\textbf{top-right}): RMSE for velocity estimates; (\textbf{bottom-left}): CDF of the position error; (\textbf{bottom-right}): CDF of the velocity error.}
    \label{fig:results_velconstrained}
\end{figure*}

\begin{figure}
    \centering
    \includegraphics[width=0.99\linewidth]{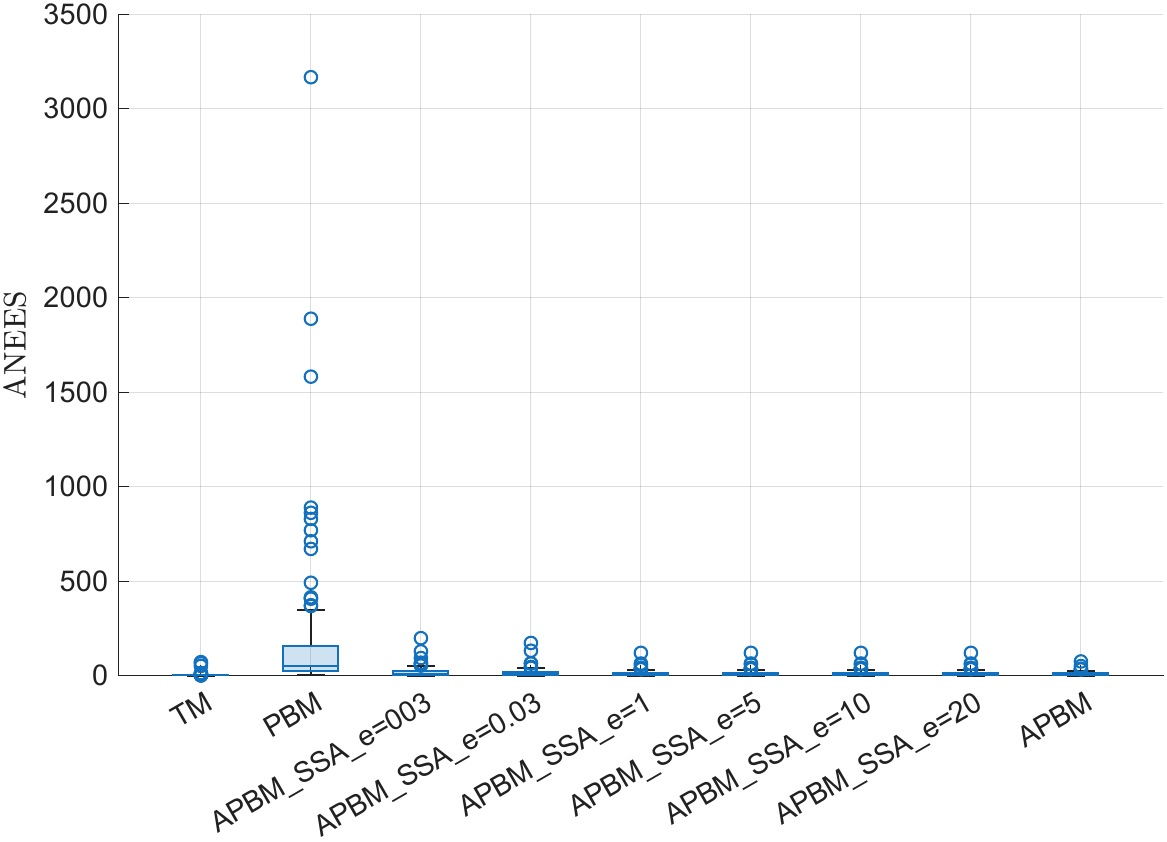}
    \caption{Results under constrained velocity - ANEEs.}
    \label{fig:ANEEs_velconstrained}
\end{figure}

We present RMSE and ANNES results using box plots computed over RMSEs and ANNESs for all time steps $k$. CDF plots were created using errors from all MC realizations and all time steps. 
The results are summarized in 
Figures~\ref{fig:results_fullyconstrained}, \ref{fig:ANEEs_fullyconstrained}, \ref{fig:results_velconstrained}, and~\ref{fig:ANEEs_velconstrained} for the different constraint setups. In the first scenario, Figures~\ref{fig:results_fullyconstrained} and~\ref{fig:ANEEs_fullyconstrained}, we applied $\rho^\mathrm{SSA}$ to all states (positions and velocities), while in 
Figures~\ref{fig:results_velconstrained} and~\ref{fig:ANEEs_velconstrained}, we applied $\rho^\mathrm{SSA}$ only to the velocities. In both scenarios, we compare the true model (TM), the \ac{pbm}, the \ac{apbm} with parameter constraint~\cite{imbiriba2022hybrid} and the $\rho^\mathrm{SSA}$-constrained APBM with different values of $\epsilon$. In all plots, the TM and \ac{pbm} are marked with dashed lines. 


Analyzing the results in Figures~\ref{fig:results_fullyconstrained} and~\ref{fig:ANEEs_fullyconstrained}, it becomes clear that \ac{apbm} leads to lower position estimation errors than \ac{pbm} while producing larger velocity estimation errors establishing our baseline. When applying the state-space constraint $\rho^\mathrm{SSA}$, $\epsilon$ controls the amount of model freedom, where smaller $\epsilon$ pushes the APBM closer to the PBM, improving velocity estimates and degrading position estimates. Conversely, by increasing $\epsilon$ the position estimates improve while velocity degrades approaching the behavior of the \ac{apbm}. This can be noticed for all metrics in Figure~\ref{fig:results_fullyconstrained}.







As discussed in Section~\ref{sec:joint_est}, one of the advantages of leveraging regularizations in the state space is that these constraints can be easily applied to only part of the states, in contrast with the regularizations applied directly over $\APBMpar$. Figures~\ref{fig:results_velconstrained} and~\ref{fig:ANEEs_velconstrained} show the results for the case where only velocities are constrained. Analyzing these results, we notice that all $\rho^\mathrm{SSA}$-constrained \ac{apbm} show comparable position estimates with the ones provided by the \ac{apbm} with velocities estimates improving as $\epsilon$ reduces. Smaller $\epsilon$ values also lead to degradation of position estimates. Although it is difficult to directly compare the results from Figures~\ref{fig:results_fullyconstrained} and~\ref{fig:results_velconstrained} since $\epsilon$ affects the regularization differently, there is some indication that constraining the whole state leads to better estimation performance. For instance, when comparing the purple line (APBM\_SSA\_e=1) in Figure~\ref{fig:results_fullyconstrained} with the golden line (APBM\_SSA\_e=0.03) in Figure~\ref{fig:results_velconstrained} we can notice similar position estimation performance but with better velocity estimation for the fully constrained scenario.
It is important to highlight that the proposed regularization approach leads to better control of the estimation performance over the different states when compared with the traditional \ac{apbm} and leads to more meaningful and interpretable ways of controlling data-driven augmentation.


\section{Conclusion}\label{sec:conc}
In this paper, we propose a new constrained state estimation approach for controlling the data-driven augmentation of adaptive state-space strategies using \acp{apbm}. The novel strategy focuses on restricting the state evolution of the augmented model to be close to the one using only the \ac{pbm}. Besides being more intuitive than the parameter regularization used in previous works, the new approach leads to better trade-offs between the estimation errors of measured and unmeasured parts of the states. Furthermore, the proposed methodology also allows for different scenarios where the constraints can be applied to only parts of the state space. Results on radar tracking problems corroborate our findings. Finally, tuning the constraint parameter $\epsilon$ is still an open problem and will be addressed in future works.

	%
	%
\addtolength{\textheight}{-6cm}   
\bibliographystyle{IEEEtran}

\end{document}